\documentstyle[12pt,draft]{article}
\newlength{\overeqskip}
\newlength{\undereqskip}
\newcommand{\nc}{\newcommand}

\newcommand{\del}{\partial}

\nc{\deldag}{\mathbin{\partial\mkern-10.5mu\big/}}
\nc{\kdag}{\mathbin{k\mkern-10mu\big/}}
\nc{\Pdag}{\mathbin{P\mkern-10mu\big/}}

\nc{\delk}{\partial_{k^3}}

\nc{\beq} {\begin{equation}}
\nc{\eeq} {\end{equation}}
\nc{\beqa}{\begin{eqnarray}}
\nc{\eeqa}{\end{eqnarray}}

\nc{\bm} {{\mathbfM}}
\nc{\bmd}{{\mathbfM^\dagger}}
\nc{\bM} {{\mathbfM}}
\nc{\bX} {{\mathbf{X}}}

\nc{\bu} {{\mathbf{u}}}
\nc{\bv} {{\mathbf{v}}}
\nc{\bud}{{\mathbf{u}^\dagger}}
\nc{\bvd}{{\mathbf{v}^\dagger}}

\nc{\nn}{\nonumber}

\begin{document}
%
%
\begin{titlepage}
\pagestyle{empty}
\baselineskip=18pt
\rightline{HD-THEP-01-23}
\rightline{NORDITA-2001-9 HE}
\rightline{May 2001}

\vskip 0.3in

\begin{center}
       {\LARGE \bf First principle derivation of semiclassical \\
                   force for electroweak baryogenesis}
\end{center}

\vskip 0.35in

\begin{center}
       {\large Kimmo Kainulainen$^\bullet$, Tomislav Prokopec$^\diamond$, \\
          Michael G. Schmidt$^\diamond$ and Steffen Weinstock$^\diamond$} \\
      \vskip 0.23in
       {\it $^\bullet$NORDITA, Blegdamsvej 17, DK-2100,
                      Copenhagen \O , Denmark} \\
      \vskip 0.1in
       {\it $^\diamond$Institute for Theoretical Physics,
                       Heidelberg University\\
                       Philosophenweg 16, D-69120 Heidelberg, Germany} \\
\end{center}

\vskip 0.35in

\centerline{\bf Abstract}
\vskip 0.3truecm

\baselineskip=18pt
\noindent

We perform a systematic gradient expansion on kinetic equations and
derive the CP-viola\-ting semiclassical force for fermions propagating
in presence of a CP-violating wall at a first order electroweak phase
transition. The force appears at order $\hbar$ in the flow term of
the kinetic equation and agrees with the semiclassical force used for
baryogenesis computations. In particular we consider the force
for charginos in both the MSSM and NMSSM. We then study the continuity
equations for the vector and axial vector currents and stress the role 
of the latter as the one containing baryogenesis sources. We also show 
that there is no CP-violating force for bosons to order $\hbar$ in 
gradient expansion.

\vskip 0.5 in

\footnoterule
\vskip 3truemm
{\small\tt
\noindent$^\bullet$kainulai@nordita.dk\\
$^\diamond$T.Prokopec@, M.G.Schmidt@, S.Weinstock@thphys.uni-heidelberg.de}

\end{titlepage}

\baselineskip=20pt

%
%

\section{Introduction}

The creation of a baryon asymmetry at a first-order electroweak phase
transition in the early universe is an attractive
proposal~\cite{KuzminRubakovShaposhnikov} because the elementary particles
and interactions involved in this process can be tested soon in accelerator
experiments. For a successful baryogenesis a coalition between CP violation,
nonequilibrium thermodynamics and baryon number violation is
needed~\cite{Sakharov}. Model calculations require a study of generation
and transport of CP-violating flows arising from interactions of fermions
with the expanding phase transition fronts. As the problem involves the
dynamics of quantum fields in a spatially varying background
it cannot be treated within the classical transport theory. While fully
general quantum Boltzmann equations can quite easily be formulated by
making use of suitably truncated Dyson-Schwinger equations for the
out-of-equilibrium two-point functions, some approximation
scheme is needed to derive a set of practically solvable, yet sufficiently
general transport equations for electroweak baryogenesis (EWBG).

A fast baryon number violating rate in the unbroken phase is a necessary
ingredient of any EWBG model. However, to avoid a wash-out of the newly
created asymmetry, the baryon number violation must turn off in the Higgs
phase.  As is well known, this is the case provided the transition is
strong enough~\cite{Shaposhnikov86}.
Since for the present experimental bounds on the Higgs mass the electroweak
phase transition in the Standard Model (SM) is not first-order
\cite{KajantieLaineRummukainenShaposhnikov}, one is lead to consider
extensions of the Standard Model. The natural candidates are 
supersymmetric models, which include the Minimal Supersymmetric Standard 
Model (MSSM)~\cite{CarenaQuirosWagnerETAL} and the nonminimal extension
(NMSSM)~\cite{PietroniDaviesFroggattMoorhouseHuberSchmidt}
with an additional Higgs-singlet field.
These models contain additional scalars which can strengthen the phase
transition as required for baryogenesis.

In supersymmetric models the bubble walls are found to be quite
slow and thick~\cite{MooreProkopec95,MorenoQuirosSeco,JohnSchmidt01,Moore00,
ClineMoore}, in the sense that $v_{\rm wall}
\ll c$ and $\ell_{\rm wall}\gg\ell_{\rm dB}$, where $\ell_{\rm dB}
\sim 1/T$ denotes a typical de Broglie wave length of thermal particles.
The latter condition is of particular importance, because it implies
that a gradient expansion in terms of $\ell_{\rm dB}/\ell_{\rm wall}$
represents a controlled, rapidly converging approximation scheme for
most of the excitations in the electroweak plasma.

In the past many heuristic attempts have been made to derive approximate
transport equations for EWBG~\cite{CohenKaplanNelson90,
NelsonKaplanCohen92,JPT-thin,JPT-thick,HuetNelson-II,Riotto-II}.
Common to all these methods
is the strategy to somehow isolate the essential quantum features of the
transport in the form of ``sources" to be inserted into classical transport
equations. Baryon production has in this way been computed in two doublet
models~\cite{NelsonKaplanCohen92,JPT-thin,JPT-thick,ClineKainulainenVischer},
MSSM~\cite{HuetNelson-II,CarenaQuirosRiottoViljaWagner,
CarenaMorenoQuirosSecoWagner,ClineJoyceKainulainen,ClineJoyceKainulainen-II,
HuberJohnSchmidt-I} and
NMSSM~\cite{HuberJohnSchmidt,HuberJohnLaineSchmidt}.
Different approaches, when applied to the same physical problem, have been
found to disagree however. In particular the sources from chargino
and squark sectors in the MSSM, found using WKB-methods
\cite{ClineJoyceKainulainen,ClineJoyceKainulainen-II,HuberJohnSchmidt},
are parametrically different from those derived by the use of the continuity
equations and the relaxation time approximation
\cite{CarenaQuirosRiottoViljaWagner,CarenaMorenoQuirosSecoWagner}, and
by other earlier attempts \cite{HuetNelson-II,HuetSather}.

In this paper we present a rigorous first-principle derivation of quantum
transport equations appropriate for baryogenesis in the limit of thick
phase boundaries ignoring collisions. We start our analysis by writing
the exact Dirac equation of motion for the dynamical Green function
(Wightman function) $G^<_{\alpha\beta}(u,v)
  \equiv i\langle \bar{\psi}_\beta(v)\psi_\alpha(u) \rangle$
with a CP-violating spatially varying pseudoscalar mass term.
For simplicity here we consider particles moving perpendicular to
the phase boundary, which effectively reduces our problem to 1+1 dimensions.
The results discussed here are not affected in any important way
when the general 3+1 dimensional case is considered~\cite{KPSW2}.
By performing a Wigner transform we obtain
a controlled expansion in gradients, or more appropriately, in
powers of $\hbar$. We show that, to first order in $\hbar$,
$G^<$ admits a spectral decomposition in
terms of on-shell quasiparticle excitations. The on-shell momenta are set
by a dispersion relation derived from the equations of motion and agree
with the results of \cite{JKP-FermionProp}, where the spectral function
${\cal A}$ was considered in gradient expansion. The on-shell distribution
functions $f_{s+}$ and $f_{s-}$ for particles and antiparticles of spin $s$,
respectively, are then shown to obey the following kinetic Liouville equations:
\begin{equation}
        \del_t f_{s\pm}
      + v_{s\pm} \del_z f_{s\pm}
      + F_{s\pm} \partial_{k_z} f_{s\pm} = 0
.
\label{firsteqn}
\end{equation}
The quantum information in~(\ref{firsteqn}) is entirely contained in the
expression for the quasiparticle energy $\omega_{s\pm}$, which shows up in the
expressions for the {\em group velocity} $v_{s\pm} \equiv k_z/\omega_{s\pm}$
and the {\em semiclassical force} $F_{s\pm} = \omega_{s\pm} dv_{s\pm}/dt$,
where $k_z$ denotes the kinetic momentum.
For example, we show that in the case of a single chiral
fermion moving in a CP-violating background with planar symmetry, represented
by a $z$-dependent complex mass $m(z) = |m(z)|\mbox{e}^{i\theta(z)}$, a
quasiparticle moving in $z$-direction with momentum $k_z$
has the energy
\beq
    \omega_{s\pm} = \omega_0 \mp \hbar \frac{s |m|^2\theta'}{2\omega^2_0}\,,
\eeq
where $\omega_0 \equiv  \sqrt{k_z^2 + |m|^2}$, and experiences the force
\beq
    F_{s\pm} = - \frac{{|m|^2}^{\,\prime} }{2\omega_{s\pm}}
            \pm \hbar \frac{s(|m|^2\theta^{\,\prime})^{\,\prime}}{2\omega_0^2}
\label{Fspm}
\,.
\eeq
We also derive explicit expressions for the
semiclassical force for a general case of $N$ mixing fermions and in
particular for charginos both in the MSSM and NMSSM. We then show
that in the case of $N$ mixing bosonic fields, such as
the squarks in the (N)MSSM, there is no CP-violating force to first
order in $\hbar$.

Our results agree with recent results obtained by the use of
the WKB-approach~\cite{ClineJoyceKainulainen-II,HuberJohnSchmidt}.
The WKB-method was originally introduced by Joyce, Prokopec and Turok
in~\cite{JPT-letter} and \cite{JPT-thick} and then applied
to the MSSM in \cite{ClineJoyceKainulainen}.
The CP-violating velocities and accelerations for fermions interacting
with a phase transition wall were correctly computed from the
WKB-dispersion relations in
Refs.~\cite{JPT-letter,JPT-thick, ClineJoyceKainulainen}.
The velocity and force in kinetic equations were obtained from
the Hamilton equations based on canonical momentum.
However, when the dispersion relation is derived by considering
the spectral function in gradient approximation~\cite{JKP-FermionProp},
the momentum appearing in the Wigner representation is the kinetic momentum.
The relevance of the kinetic momentum as the true physical variable
in the WKB-picture was first realized by Cline, Joyce and
Kainulainen~\cite{ClineJoyceKainulainen-II}, who also showed how
it can be consistently incorporated into kinetic theory
leading to equations identical to~(\ref{firsteqn}-\ref{Fspm}).

The outstanding contribution of the present work is a controlled first
principle derivation of the kinetic equation~(\ref{firsteqn}). This is
important because of a considerable controversy in literature concerning 
the transport equations to be used for EWBG calculations.
Moreover, our treatment in principle allows a study of the plasma dynamics
beyond first order in $\hbar$, which cannot be achieved by WKB-methods.
As an example, at second order in $\hbar$ the full equations do not admit
the spectral decomposition solution for $G^<$; for a scalar field
this problem is considered in~\cite{BKP}.

Let us mention that in a related work \cite{VasakGyulassyElze}
the Liouville equations for fermions
in presence of a classical gauge field have been considered in gradient
approximation. The crucial
role of the constraint equations in the derivation of the kinetic
equations was then stressed in~\cite{ZhuangHeinz-II,ZhuangHeinz-I}.
The problem of a pseudo-scalar mass term in
kinetic equations has been considered in
Refs.~\cite{ShinRafelski,ZhuangHeinz-II}, but these
authors discussed the flow term only to zeroth (classical) order in
$\hbar$, whereas the spin dependent force essential for EWBG arises
only at quantum level, as we show here.

Inclusion of interaction terms gives rise to yet another source which is
of first order in $\hbar$, namely the {\em spontaneous baryogenesis}
(SBG) source of Ref.~\cite{CohenKaplanNelson91}. The SBG source appears
because the CP-violating split in the dispersion relation causes the
CP-conjugate states to relax towards different local equilibria in the
bubble wall. Thus a first principle derivation of the SBG source requires
not only a consistent expansion in $\hbar$, which we have done, but also
a consistent expansion in relevant coupling constants. The latter is
necessarily a model dependent problem and shall be considered 
elsewhere. However, to facilitate comparison 
with literature we derive the vector current to order $\hbar$ in gradient 
expansion which displays the SBG source in the
relaxation time approximation. When applied to the MSSM, our results
differ from Refs.~\cite{HuetNelson-I}, \cite{Riotto-II}
and~\cite{CarenaMorenoQuirosSecoWagner}.

The paper is organized as follows. In section 2 we derive the
Liouville equations for a single Dirac fermion with a spatially varying
complex mass term.  In section 3 we generalize these results to
the case of $N$ mixing fermionic fields and study the case
of mixing charginos in both the MSSM and NMSSM. We then in section~4
consider the case of $N$ mixing scalar fields and show that there is
no CP-violating source to first order in $\hbar$.  In section 5 we
study the continuity equations for both vector and axial vector current,
and spontaneous baryogenesis in the relaxation time approximation,
and make a comparison with literature. For example, in contrast to what
is claimed in~\cite{Riotto-I} and \cite{Riotto-II}, we find that the
continuity equation for the vector current contains no CP-violating source
in the absence of collisions. On the other hand, in the continuity equation
for the axial current there are CP-violating sources that can be related
to higher moments expansion of the semiclassical Boltzmann equation.
Finally, section 6 contains a discussion and summary.

\section{Fermionic field with a complex mass}

We first consider the dynamics of a fermionic field with a complex
spatially varying mass term. More precisely, we take our system to be
described by the effective lagrangian
\beq
      {\cal L} = i\bar{\psi}\deldag\psi - \bar{\psi}_Lm\psi_R
            - \bar{\psi}_Rm^*\psi_L + {\cal L}_{\rm int}
\,,
\label{lagrangian}
\eeq
where ${\cal L}_{\rm int}$ contains interactions and
\beq
      m(x) = m_R(x) + i m_I(x) = |m(x)|\mbox{e}^{i\theta(x)}
\label{mass1}
\eeq
is a space-time dependent mass term arising from an interaction with some
CP-violating scalar field condensate. We are primarily interested in the case
where $m$ arises from the Higgs field condensate at a first order electroweak
phase transition. As the bubbles of broken phase grow several orders of
magnitude larger than the wall width before coalescence, the wall can
be approximated by a planar interface to good accuracy. We
therefore consider a mass term in the bubble wall frame which is
only a function of the spatial coordinate orthogonal to the wall,
$m= m(z)$.

Our focus in this paper is on the semiclassical treatment of fermions in
presence of background fields.  In particular we derive the flow term 
for a fermionic kinetic equation with a nontrivial force induced by
the CP-violating mass parameter (\ref{mass1}). To keep the discussion simple,
we relegate the explicit treatment of collision terms and self-energy
corrections which are induced by the specific interactions included in
${\cal L}_{\rm int}$, to later publications~\cite{KPSW2}. Moreover, we are 
interested in the case of wide walls, so that the de Broglie wave length 
$\ell_{\rm dB}$ of a typical excitation is small in comparison 
with the wall width, $\ell_{\rm dB} \ll \ell_{\rm wall}$.
This condition is amply satisfied at the electroweak phase transition,
where typically $\ell_{\rm dB} \sim 1/T$ and
$\ell_{\rm wall} \sim 10/T$~\cite{MorenoQuirosSeco}.

With the above assumptions we now develop the equations of motion
for the Wightman function
\beq G^<_{\alpha\beta}(u,v) \equiv
       i\left< \bar{\psi}_\beta(v)\psi_\alpha(u) \right>
\label{G-less}
\eeq
in a consistent expansion in gradients of the background fields. Here
$\left<\cdot \right>$ denotes the expectation value with respect to the
initial state. The function $G^<$ describes the statistical properties of
an out-of-equilibrium system. It corresponds to the off-diagonal part of 
the fermionic two-point function in the
Schwinger-Keldysh formalism~\cite{SchwingerKeldysh},
which indeed is the method of choice to derive the equations of motion
for $G^<$ including interactions. However, in their absence all one needs
is the familiar Dirac equation. Dropping interactions we have from
Eq.~(\ref{lagrangian}):
\beq
     \left( i\deldag_u \! - \, m_R(u) - i\gamma^5m_I(u) \right) \psi(u) = 0 \,.
\label{Dirac}
\eeq
Multiplying (\ref{Dirac}) from the left by the spinor $i\bar{\psi}(v)$ and
taking the expectation value one finds:
\beq
\left(i\deldag_{u}\! - \,  m_R(u) - i\gamma^5m_I(u) \right) G^<(u,v) = 0
\, .
\label{F}
\eeq
This equation and the hermiticity property
\beq
\left[i\gamma^0G^<(u,v)\right]^\dagger =i\gamma^0G^<(v,u),
\label{hermiticity}
\eeq
which can be immediately inferred from the definition~(\ref{G-less}),
completely specifies $G^<$.

In order to study equation~(\ref{F}) in gradient expansion we perform a
Wigner transform of $G^<$ to the mixed representation, {\em i.e.\ } a
Fourier transform with respect to the relative coordinate $r\equiv u-v$:
\beq
G^<(x,k) \equiv \int d^{\,4} r \, e^{ik\cdot r} G^<(x + r/2,x-r/2),
\label{wigner1}
\eeq
where $x=(u+v)/2$ denotes the center-of-mass coordinate. The crucial
advantage of the representation (\ref{wigner1}) is that it separates the
internal fluctuation scales, described by momenta $k$, from the external
ones which show up as a dependence of $G^<$ on $x$, and thus gives
us the chance of exploiting possible hierarchies between these scales.
In the Wigner representation equation~(\ref{F}) becomes
\beq
\left(\hat{\kdag} - {\hat m}_0(x) -i {\hat m}_5(x) \gamma^5\right) \,
G^< = 0,
\label{G-less-eom}
\eeq
where we use the following convenient shorthand notation
\beqa
{\hat k}_\mu &\equiv& k_\mu + \frac{i}{2}\partial_\mu
\label{shorts-k} \\
{\hat m}_{0(5)} &\equiv& m_{R(I)}
   e^{-\frac{i}{2}\stackrel{\!\!\leftarrow}{\partial_x}\cdot\,\partial_k}.
\label{shorts}
\eeqa
The original local equation (\ref{F}) for $G^<$ is thus transformed into
an equation involving an infinite series in gradients. This in
fact can be viewed as an expansion in powers of the Planck constant $\hbar$.
We have set $\hbar \rightarrow 1$, but a dimensionful $\hbar$ can at any
stage be easily restored by the simple replacements $\partial_x
\rightarrow \hbar\,\partial_x$ and $G^< \rightarrow \hbar^{-1}\,G^<$.

Because of the planar symmetry (here we do not consider initial states
of the plasma that break this symmetry) $G^<$ can depend
only on the spatial coordinate orthogonal to the wall, $z\equiv x^3$.
We also consider equation (\ref{G-less-eom}) only in a frame where
the momentum parallel to the wall vanishes, $\vec k_\parallel = 0$.  This
last assumption effectively reduces the problem to 1+1 dimensions, and we
can cast equation (\ref{G-less-eom}) into the form
\beq
\left({\hat k}_0 + {\hat k}_z \gamma^0 \gamma^3
          - {\hat m}_0\gamma^0 + i {\hat m}_5 \gamma^0\gamma^5 \right)
           i \gamma^0G^< = 0 ,
\label{G-less-eom2}
\eeq
where ${\hat k}_0 = k_0 + \frac{i}{2}\partial_t$ and
${\hat k}_z = k_z - \frac{i}{2}\partial_z$. The differential
operator  in (\ref{G-less-eom2}) is entirely spanned by
a closed 1+1-dimensional subalgebra of the full 3+1-dimensional
Clifford algebra. Moreover, it commutes with the operator
$S^3 = \gamma^0\gamma^3\gamma^5$, which measures spin $s$
in $z$-direction. $s$ is thus a good quantum number in the
frame $\vec k_\parallel = 0$, which motivates to seek solutions 
for $i\gamma^0G^<$ which satisfy 
$S^3i\gamma^0G^<_s  = i\gamma^0G^<_s S^3 = si\gamma^0G^<_s$.
Working in a convenient chiral representation this condition 
leads immediately to the following spinor structure:
\beq
-i\gamma^0G^<_s =   \frac12(1+s\sigma^3)\otimes g^<_s,
\label{connection}
\eeq
where $\sigma^3$ is the Pauli matrix referring to spin in 
$z$-direction. $g^<_s$ has indices in the remaining two dimensional 
chiral space and can also be written in terms of the Pauli 
matrices $\rho^i$ as follows:
\beq
g^<_s \equiv \frac12 \left( g^{s}_0 +  g^{s}_i \rho^i \right) .
\label{glesss}
\eeq
The decomposition~(\ref{connection}) contains the implicit assumption that
$i\gamma^0G^<$ does not mix spins. We expect that this is
a good approximation in the electroweak plasma.
The signs and normalizations in (\ref{connection}-\ref{glesss}) are
chosen such that $g_0^{s}$ measures the number density of particles with spin
$s$ in phase space. With these simplifications we can now reduce our
original $4\times4$ problem to a two-dimensional one by effecting the
replacements
\beq
              \gamma^0 \; \rightarrow \; \rho^1,  \qquad
-i\gamma^0\gamma^5 \; \rightarrow \; \rho^2,  \qquad
             -\gamma^5 \; \rightarrow \; \rho^3
\label{subs}
\eeq
and we find
\beq
\left( {\hat k}_0 - s {\hat k}_z\rho^3
          - \rho^1 {\hat m}_0 - \rho^2 {\hat m}_5  \right) g^<_s = 0.
\label{Gs-eom}
\eeq
A similar procedure is used in~\cite{JKP-FermionProp} for a treatment
of the fermionic propagator. Equation~(\ref{Gs-eom}) may look simple, but
it still constitutes
a set of four coupled complex (or eight coupled real) differential
equations, which we shall now analyze.  We first find the four independent
complex equations for $g^{s}_a$ ($a=0,i$) by multiplying (\ref{Gs-eom})
successively by $1$ and $\rho^i$, and taking the trace:
\def\hi{\phantom{i}}
\beqa
{\hat k}_0 g^{s}_0  - \hi s{\hat k}_z g^{s}_3
            - \hi {\hat m}_0 g^{s}_1 - \hi {\hat m}_5 g^{s}_2 &=& 0
\label{eq-sigma1} \\
{\hat k}_0 g^{s}_3  - \hi s{\hat k}_z g^{s}_0
            - i   {\hat m}_0 g^{s}_2 + i   {\hat m}_5 g^{s}_1 &=& 0
\label{eq-sigma2} \\
{\hat k}_0 g^{s}_1  + i   s{\hat k}_z g^{s}_2
            - \hi {\hat m}_0 g^{s}_0 - i   {\hat m}_5 g^{s}_3 &=& 0
\label{eq-sigma3} \\
{\hat k}_0 g^{s}_2  - i   s{\hat k}_z g^{s}_1
            + i   {\hat m}_0 g^{s}_3 - \hi {\hat m}_5 g^{s}_0 &=& 0.
\label{eq-sigma4}
\eeqa
As a consequence of Eq.~(\ref{hermiticity}) the matrices $g^<_{s}$ are
hermitean so that $g_a^{\,s}$ are real functions. We then have twice
as many equations as independent functions corresponding to real and
imaginary parts of Eqs.~(\ref{eq-sigma1}-\ref{eq-sigma4}), and hence one half
of the equations must correspond to the constraints on the solutions
of the other half; those equations are kinetic equations.
This was first pointed out in the context of
kinetics of fermions by Zhuang and Heinz~\cite{ZhuangHeinz-II}.
As one sees from~(\ref{shorts-k}), no time derivatives appear in the real
parts of~(\ref{eq-sigma1}-\ref{eq-sigma4}) so that they indeed provide four
{\em constraint equations} (CE) on the solutions of four
{\em kinetic equations} (KE). These contain time derivatives
and correspond to the imaginary parts of~(\ref{eq-sigma1}-\ref{eq-sigma4}).

Because we have put no restrictions to the form of the $\hat m$-operators,
Eqs.~(\ref{eq-sigma1}-\ref{eq-sigma4}) are still valid to any order in
gradients in the frame where $\vec k_\parallel=0$. In what follows we
assume that the mass is a slowly varying function of $x$ and truncate
gradient expansion at second order, which is the lowest order at which
CP-violating effects can be discussed consistently.
This method is not adequate for problems involving quantum mechanical 
reflection, which require nonperturbative treatment in $\hbar$. 
With the truncation, ${\hat m}_{0(5)}$ 
in~(\ref{eq-sigma1}-\ref{eq-sigma4}) simplifies to
\beq
{\hat m}_{0(5)} \simeq m_{R(I)} + \frac{i}{2} m_{R(I)}' \partial_{k_z}
                                    - \frac{1}{8} m_{R(I)}''\partial^2_{k_z}.
\label{shorts2c}
\eeq
Even with this truncation, we are facing a problem involving eight coupled
second order partial differential equations. Our next task is to reduce these
to a single equation governing the dynamics of the fermionic
two-point function.

\subsection{Constraint equations}

Let us first consider the constraint equations. They consist of four
homogeneous equations for four functions, which implies that there is one
constraint which gives rise to the dispersion relation.  While this property
remains true to any order in gradients (or equivalently in $\hbar$), we only
need to work to first order to find the nontrivial result we are looking for.
To this order we have
\beqa
&& k_0 g^{s}_0 - sk_z g^{s}_3 - m_R g^{s}_1 - m_I g^{s}_2 = 0
\label{CEA} \\
&& k_0 g^{s}_3 - sk_z g^{s}_0 + \frac12 m_R' \partial_{k_z} g^{s}_2
                          - \frac12 m_I' \partial_{k_z} g^{s}_1 = 0
\label{CEB} \\
&& k_0 g^{s}_1 + \frac{s}{2} \partial_z g^{s}_2
               - m_R g^{s}_0 + \frac12 m_I' \partial_{k_z} g^{s}_3 = 0
\label{CEC} \\
&& k_0 g^{s}_2 - \frac{s}{2} \partial_z g^{s}_1
               - m_I g^{s}_0 - \frac12 m_R' \partial_{k_z} g^{s}_3 = 0.
\label{CED}
\eeqa
We first use the constraint equations~(\ref{CEB}-\ref{CED})
iteratively up to first order to express $g^{s}_1$, $g^{s}_2$ and
$g^{s}_3$ in terms of $g^{s}_0$ and $\partial_{k_z} g^{s}_0$, and then
insert the results into~(\ref{CEA}). Remarkably all terms proportional
to $\partial_{k_z} g^{s}_0$ cancel and we find that to first order
in gradients $g^{s}_0$ satisfies the algebraic equation
\beq
\left( k_0^2 -  k_z^2 - |m|^2  + \frac{s}{k_0}|m|^2\theta^{\,\prime}
       \right) g^{s}_0 = 0
\label{constraint-g0}
\,.
\eeq
This admits the spectral solution $g^{s}_0 = \pi n_s |k_0| 
\delta(\Omega_s^2)$, which can also be written as
\beqa
g^{s}_0  &=& \sum_\pm \frac{\pi }{2 Z_{s\pm}}
                 \, n_s \,\delta(k_0 \mp \omega_{s\pm})\,.
\label{spectral-dec}
\eeqa
Here $n_s(k_0,k_z,z)$ are nonsingular
functions which are, as we show below, related to the on-shell distribution
functions. The indices $\pm$ refer to the sign of $k_0$, to be eventually
related to particles and antiparticles. The energy $\omega_{s\pm}$ is
specified by the roots of the equation
\beq
\Omega^2_s \; \equiv \;
       k_0^2 - k_z^2 - |m|^2 + \frac{s}{k_0}|m|^2\theta^{\,\prime} \; = \; 0,
\label{DR1}
\eeq
and the normalization factor $Z_{s\pm}$ is defined as
\beq
     Z_{s\pm} \equiv \frac{1}{2 \omega_{s\pm}\,} |\partial_{k_0}
                      \Omega^2_s|_{k_0=\pm\omega_{s\pm}}.
\label{normfac}
\eeq
To first order in gradients these can be solved iteratively:
\beqa
     \omega_{s\pm} &=& \omega_0 \mp s \frac{|m|^2\theta'}{2\omega^2_0}\,,
     \qquad \qquad \omega_0 = \sqrt{ k_z^2 + |m|^2}
\label{dispersion1} \\
      Z_{s\pm} &=&  1 \mp s\frac{|m|^2\theta'}{2\omega^3_0}
      = \frac{\omega_{s\pm}}{\omega_0}
\,.
\label{normfac2}
\eeqa
Equation (\ref{dispersion1}) defines the physical dispersion relations for
particles and antiparticles of a given spin $s$. Due to the derivative
corrections the spin degeneracy is lifted at first order in gradients and
hence particles (antiparticles) of different spin experience different 
accelerations in a spatially varying background, as we shall see in more 
detail below.

Solution (\ref{spectral-dec}) nicely illustrates how the constraint
equations operate.  Solving (\ref{CEA}-\ref{CED}) consistently to first
order accuracy constrains the solutions of the kinetic equation to sharp, 
locally varying energy shells given by~(\ref{dispersion1}). 
One way of understanding this is as follows. The CP-violating phase 
$\theta'$ in (\ref{constraint-g0}) can be related to an axial gauge 
field~\cite{JPT-letter,JPT-thick} which, to leading order in gradients, 
lifts the degeneracy in the dispersion relation, but does not spoil the 
quasiparticle picture, just as it is the case with a vector gauge field.
We should note 
however, that the confinement to sharp energy shells does not persist beyond 
first order in gradients. While for noninteracting fermions one can always 
express $g^s_{1,2,3}$ in terms of $g^s_0$, at higher orders more complicated
derivative structures arise, as the constraint cannot be written as a
simple algebraic equation with a spectral solution. For a treatment of
such a situation in the case of a scalar field see~\cite{BKP}.

Eq.~(\ref{dispersion1}) is identical with the results derived  earlier by
WKB-methods in \cite{ClineJoyceKainulainen-II} and simultaneously via
the field-theoretic technique of spectral integrals
in~\cite{JKP-FermionProp}. From the
WKB-point of view the present derivation comes as a welcome verification of
the result obtained by an intuitive, but less fundamental approach. The
agreement with the field theoretical calculation of \cite{JKP-FermionProp} on
the other hand was to be expected.  In~\cite{JKP-FermionProp} it was shown that
the integral over the spectral function, 
defined as a difference of the retarded and advanced
Green functions ${\cal A} = (i/2)(G_{\rm ret}-G_{\rm adv})$, projects
test functions onto energy shells (\ref{dispersion1}); however,
in the collisionless limit ${\cal A}$ satisfies the same equation of
motion~(\ref{F}) as the Wightman function $G^<$. The trace of
$\gamma^0{\cal A}$ in particular satisfies equation~(\ref{constraint-g0}),
and can be obtained from~(\ref{spectral-dec}) by the replacement
$n_s \rightarrow 1$. We can hence immediately check the sum-rule to
the accuracy at which we are working. Indeed,
\beqa
       \int_{-\infty}^\infty \frac{d k_0}{\pi } {\rm Tr}\,\gamma^0{\cal A}_s
        \, = \,  \sum_\pm
        \int_{-\infty}^\infty \frac{d k_0 }{2Z_{s\pm} }
        \delta(k_0 \mp \omega_{s\pm})  \, = \, 1.
\label{sumrule}
\eeqa

We finally note that equation (\ref{DR1}) has additional poles at
$k_0 \simeq s|m|^2\theta'/2\omega_0$, which we have left out in the
decomposition (\ref{spectral-dec}). These poles correspond to unphysical,
but harmless, tachyonic modes which arise only because our solutions for
the constraint equations involve an expansion in inverse powers of
$k_0$, which breaks down already for $k_0$ much larger than the
value associated with these poles. Note that their contribution to
the sum rule~(\ref{sumrule}) vanishes when summed over spins.

\subsection{Kinetic equations}

We now turn our attention to the kinetic equations. We are primarily
interested in the equation for $g^{s}_0$ which carries information 
on the particle density in phase space.  From (\ref{eq-sigma1}) we 
have
\beq
      \partial_t g^{s}_0 + s\partial_z g^{s}_3
            - m_R' \partial_{k_z}g^{s}_1
            - m_I' \partial_{k_z}g^{s}_2 = 0,
\label{KEA}
\eeq
which is correct up to second order in gradients (first order in 
$\hbar$). Just as in the previous section we use the constraint
equations~(\ref{CEB}-\ref{CED}) to express $g^{s}_{1}$, $g^{s}_{2}$ 
and $g^{s}_{3}$ in terms of $g_0^s$ and arrive at an equation for 
$g^{s}_0$ alone. To second order in gradients (first order in $\hbar$) 
it reads
\beq
         k_0 \del_t  g^{s}_0 + k_z\partial_z g^{s}_0
       - \left(   \frac{1}{2} {|m|^2}^{\,\prime}
                - \frac{s}{2k_0}(|m|^2\theta^{\,\prime})^{\,\prime}
                         \right)  \partial_{k_z}g^{s}_0 = 0.
\label{boltzmann1}
\eeq
We have so far used three out of four constraint equations.  To find
the acceptable solutions satisfying all constraints, we must yet impose
the restriction onto the functional space spanned by (\ref{spectral-dec}).
Because of the $\delta$-function in the decomposition (\ref{spectral-dec})
this is of course trivial. All we need to do is to insert
(\ref{spectral-dec}) into (\ref{boltzmann1}) and integrate over
the positive and negative frequencies $k_0$. We then get the
following form for the Liouville equation
\beq
        \del_t f_{s\pm}
      + v_{s\pm} \del_z f_{s\pm}
      + F_{s\pm} \partial_{k_z} f_{s\pm} = 0,
\label{boltzmann1-sc}
\eeq
where
\beqa
     f_{s+} &\equiv& n_s(\omega_{s+},k_z,z) \nn \\
     f_{s-} &\equiv& 1 - n_s(-\omega_{s-},-k_z,z)
\label{fs}
\eeqa
are the distribution functions for particles and antiparticles with
spin $s$, respectively. These definitions are motivated by the equilibrium
result $n_s^{\rm eq} = 1/(e^{\beta k_0}+1)$, where $\beta=1/T$ is the inverse
temperature. The quasiparticle {\em group velocity} $v_{s\pm}$ appearing in
Eq.~(\ref{boltzmann1-sc})
is given by
\beq
v_{s\pm} = \frac{k_z}{\omega_{s\pm}}
,
\label{vel}
\eeq
where $k_z$ is the kinetic momentum. The spin-dependent and CP-violating 
{\em semiclassical force} reads
\beq
F_{s\pm}  =  - \frac{{|m|^2}^{\,\prime} }{2\omega_{s\pm}}
                      \pm \frac{s(|m|^2\theta')'}{2\omega_0^2}
.
\label{scforce1}
\eeq
Equations~(\ref{boltzmann1-sc}-\ref{scforce1}) are among the main results
of this paper. Incidentally, the form (\ref{scforce1}) for the semiclassical
force $F_{s\pm}=\omega_{s\pm}dv_{s\pm}/dt$ was already found by Joyce, 
Prokopec and Turok~\cite{JPT-letter}. To obtain kinetic equations by 
WKB-methods the authors of \cite{JPT-thick} used canonical variables 
however. The resulting equations are not invariant under reparametrization 
of the wave functions, and hence care is required when specifying local 
thermal equilibrium in derivation of transport equations relevant for 
baryogenesis. Cline, Joyce and Kainulainen~\cite{ClineJoyceKainulainen-II} 
introduced the kinetic momentum as a physical 
variable in the kinetic equations and obtained the unique reparametrization 
invariant transport equations identical with~(\ref{boltzmann1-sc}) and 
(\ref{vel}-\ref{scforce1}). The outstanding contribution of the present 
work is in a controlled first principle derivation of these equations 
without any {\em a priori} assumptions.

Let us finally note that equation (\ref{boltzmann1}) could have been
obtained by taking the bilinear $\Diamond$-derivative of the constraint
equation (\ref{constraint-g0}),  where the $\Diamond$-derivative is
defined by
\beq
\Diamond\{a\}\{b\} \equiv \frac 12\left(
\partial_t a\;\partial_{k_0} b
-\partial_z a\;\partial_{k_z} b
-\partial_{k_0}a\; \partial_t b
+ \partial_{k_z} a\;\partial_z b
\right).
\label{Diamond}
\eeq
This is no coincidence, and even more generally, in the collisionless limit
the kinetic equation can be obtained by effecting the
$\tan \Diamond$-derivative on the constraint equation.

\subsection{Currents}

It is instructive to study the expressions for physical currents in order
to shed light on various functions we have encountered in our derivation.
Of particular relevance for baryogenesis are the vector and axial vector
currents. By making use of~(\ref{G-less})
and (\ref{connection}-\ref{glesss}) one finds
\begin{eqnarray}
       j^\mu &\equiv&  \left<\bar{\psi}(x)\gamma^\mu\psi(x)\right>
     = \sum_{s=\pm 1}\int \frac{d^{\,2}k}{(2\pi)^2}
            \left(\, g^s_0, sg^s_3 \, \right)
\nonumber\\
      j^\mu_5 &\equiv& \left<\bar{\psi}(x)\gamma^\mu\gamma^5\psi(x)\right>
     = \sum_{s=\pm 1}\int \frac{d^{\,2}k}{(2\pi)^2}
            \left(\, g^s_3, sg^s_0 \, \right)
,
\label{vector-axial}
\end{eqnarray}
where we have restricted ourselves to 1+1-dimensions so that
$d^2k = dk_zd k_0$. This shows that $g^s_0$ is the usual number
density in phase space, whereas $g_3^s$ represents the axial charge density.
An important consequence of the constraint equations is that
there is only one independent dynamical function, here chosen to be $g^s_0$,
while all others can be related to $g^s_0$ via the constraint equations
(\ref{CEB}-\ref{CED}). In particular, $g_3^s$ can be written as
\beq
    g_3^{\,s} = \Big(
      s\frac{k_z}{k_0} + \frac{1}{2k_0^2} |m|^2\theta '\partial_{k_z}
                \Big) \, g_0^{s}  \,.
\label{g3-g0}
\eeq
The nontrivial gradient correction appears as a total derivative and hence
vanishes upon the $k_z$-integration. Using the decomposition
(\ref{spectral-dec}) one finds
\beqa
     j^\mu_{s\pm}  &=&
               \phantom{s}
          \int \frac{dk_z }{8\pi Z_{s\pm}} \, (1,v_{s\pm}) \, f_{s\pm}
       \;  =   \phantom{s}
          \sum_{s_{k_z}=\pm} s_{k_z}
          \int \frac{d\omega }{8\pi}\,
          (\frac{1}{v_{s\pm}},1) \, f_{s\pm}
\label{vector2}
\\
     j^\mu_{5s\pm} &=&
          s \int \frac{dk_z }{8\pi Z_{s\pm}} \, (v_{s\pm},1) \, f_{s\pm}
       \;  =
          s \sum_{s_{k_z}=\pm}
          s_{k_z} \int \frac{d\omega }{8\pi}
          \, (1,\frac{1}{v_{s\pm}}) \, f_{s\pm}
       \, ,
\label{axial2}
\eeqa
where $s_{k_z}$ denotes the sign of $k_z$, and we discarded the vacuum
contribution. In the last step we used $\partial_{k_z}\omega_{s\pm} =
Z^{-1}_{s\pm}k_z/\omega_{s\pm}$. The lower limit in the $\omega$-integrals
is $|m| \mp s\theta'/2$. The functions $f_{s\pm}$ are the correctly 
normalized distribution functions, and they retain the correct 
physical interpretation in that the particle flux is not affected 
by CP-violating effects, while the density may be 
either enhanced or suppressed, as given by the inverse velocity. 
The current (\ref{vector2}) was computed by WKB-methods
in \cite{ClineJoyceKainulainen-II}. The correct result for $j^0_{s\pm}$
was also found in \cite{JKP-FermionProp} by field-theoretical methods.
However, $j^3_{s\pm}$ found in \cite{JKP-FermionProp} does not
agree with (\ref{vector2}) because the spinor structure used for
$i\gamma^0G^<$ was too simple.

\subsection{Interactions}

In all discussions above, we have left out the effects of interactions.
This was done to avoid the necessity to use the full machinery of the
Schwinger-Keldysh formalism~\cite{SchwingerKeldysh}, and to keep things
as simple as possible. Moreover, unlike the treatment of the flow term
of the kinetic
equation presented above, including interactions is necessarily a model
dependent task. Nevertheless it is a simple matter to write a {\em formally
exact} equation of motion for $G^<$ including the collision terms. Instead
of~(\ref{F}) we then have
\beq
(i \deldag - m_R - im_I\gamma^5) \, iG^< - \Sigma_{\rm R} \odot iG^<
            \; = \; \Sigma^< \odot G_R
              + \frac12 \Big( \Sigma^> \odot G^< - G^< \odot \Sigma^> \Big)
,
\label{KEfull}
\eeq
where $A \odot B (u,v) \equiv \int dw \; A (u,w) B(w,v)$. 
$G_R$ is the real part of the (retarded) propagator, and 
the function $\Sigma_{\rm R}$ contains the real part of the self-energy
corrections including the singular (tadpole) interactions which can be 
resummed to a renormalized mass term. The terms in parentheses give rise 
to the usual collision term, where the self-energies $\Sigma^{<,>}$ arise 
only from nonlocal loop contributions to the Dyson-Schwinger equations.
However, as mentioned above, the exact form of these terms depends on the
theory considered. While their treatment is not conceptually difficult,
their inclusion brings a considerable amount of technical complications.
We shall consider the problem of including collisions 
elsewhere~\cite{KPSW2}.

\section{Mixing fermionic fields}

In practical applications, such as in supersymmetric models, one needs to
consider cases where several fermion flavours are mixed by a spatially
varying mass matrix.  We therefore consider a theory with the mass
lagrangian
\beq
     {\cal L}_{\rm mass} \, = \,
          - \bar{\psi}_L M \psi_R  - \bar{\psi}_R M^\dagger \psi_L
\,,
\label{Nlagrangian}
\eeq
where $M$ is a complex (in general nonhermitean) $N\times N$ matrix
with spatially varying components.
We denote the flavour degree of freedom by an additional index $i$ to
the spinor $\psi_{\alpha,i}(x)$, so the Wightman function becomes a matrix
in the product space of spinor and flavour:
\beq
     G^<_{\alpha\beta,ij}(u,v)
      = i \left< \bar{\psi}_{\beta,j}(v) \psi_{\alpha,i}(u) \right>
\,.
\label{multigreen}
\eeq
The flavour degree of freedom plays no role in the derivation of the
equations of motion for $G^<_{\alpha\beta,ij}$ in the steps analogous
to going from (\ref{F}) to~(\ref{eq-sigma1}-\ref{eq-sigma4}) in
the single fermion field case, because those steps dealt
only with the spinor structure of $G^<$.  We can thus
immediately write an equation analogous to (\ref{Gs-eom}):
\beq
\left( {\hat k}_0 - s {\hat k}_z\rho^3
          - \rho^1 {\hat M}_0 - \rho^2 {\hat M}_5  \right) g^<_s = 0.
\label{Gs-eomN}
\eeq
The sole, but significant difference to (\ref{Gs-eom}) is that
$g_{s}^<$ is now an $N\times N$-matrix in the flavour space, and the
mass terms have become $N\times N$-matrix operators
\beqa
{\hat M}_0 \! &=& \! \phantom{-} \frac12 ( \hat M + \hat M^\dagger) \\
{\hat M}_5 \! &=& \! - \frac{i}{2} ( \hat M - \hat M^\dagger)
\label{shorts2d}
\eeqa
where $\hat M \equiv
M e^{\frac{i}{2} \stackrel{\!\!\leftarrow}{\partial_z} \partial_{k_z}}$
and $\hat M^\dagger \equiv  M^\dagger
e^{\frac{i}{2} \stackrel{\!\!\leftarrow}{\partial_z} \partial_{k_z}}$.
The extra flavour structure of course complicates the solution of equation
(\ref{Gs-eomN}) and it turns out to be convenient to perform a rotation to 
the basis where the lowest order mass matrix is diagonal.  Because $M$ in 
general can be nonhermitean, the diagonalization requires a 
{\em biunitary} transformation
\beq
        U M  V^\dagger = M_d \,,
\label{rotation}
\eeq
where $U$ and $V$ are the unitary matrices which diagonalize the
hermitean matrices $MM^\dagger$ and $M^\dagger M$, respectively. After
the rotation we can write (\ref{Gs-eomN}) in the component form in the
diagonal basis as
\def\hi{\phantom{i}}
\beqa
          (k_0 + \frac{i}{2} {\cal D}^-_t) g^s_{0d}
- \hi s (k_z - \frac{i}{2} {\cal D}^-_z) g^s_{3d}
            - \hi  {\hat M}_{0d} g^s_{1d}
            - \hi  {\hat M}_{5d} g^s_{2d} &=& 0
\label{Neq-sigma1} \\
          (k_0 + \frac{i}{2} {\cal D}^-_t) g^s_{3d}
- \hi s (k_z - \frac{i}{2} {\cal D}^-_z) g^s_{0d}
            - i    {\hat M}_{0d} g^s_{2d}
            + i    {\hat M}_{5d} g^s_{1d} &=& 0
\label{Neq-sigma2} \\
          (k_0 + \frac{i}{2} {\cal D}^+_t) g^s_{1d}
   + i  s (k_z - \frac{i}{2} {\cal D}^+_z) g^s_{2d}
            - \hi  {\hat M}_{0d} g^s_{0d}
            - i    {\hat M}_{5d} g^s_{3d} &=& 0
\label{Neq-sigma3} \\
          (k_0 + \frac{i}{2} {\cal D}^+_t) g^s_{2d}
   - i  s (k_z - \frac{i}{2} {\cal D}^+_z) g^{s}_{1d}
            + i    {\hat M}_{0d} g^s_{3d}
            - \hi  {\hat M}_{5d} g^s_{0d} &=& 0.
\label{Neq-sigma4}
\eeqa
The `covariant derivatives' appearing in
(\ref{Neq-sigma1}-\ref{Neq-sigma4}) are defined as
\beqa
{\cal D}^\pm_t &\equiv& \partial_t -i [\Sigma_t, \, \cdot \,]_-
                                   -i s [\Delta_z, \, \cdot \,]_\pm
\\
{\cal D}^\pm_z &\equiv& \partial_z -i [\Sigma_z, \, \cdot \,]_-
                                   -i s [\Delta_t, \, \cdot \,]_\pm
\label{covderivs}
\eeqa
where the brackets $[\cdot,\cdot ]_-$ refer to commutators and
$[\cdot,\cdot]_+$ to anticommutators and
\beqa
       \Sigma_\mu    &\equiv& \frac{i}{2}
        \big( V \del_\mu V^\dagger + U \del_\mu U^\dagger \big)
\\
      \Delta_\mu &\equiv& \frac{i}{2}
        \big( V \del_\mu V^\dagger - U \del_\mu U^\dagger \big) \,.
\label{Isodelta}
\eeqa
It should be noted that, while the relation (\ref{rotation}) allows
arbitrary phase redefinitions $U \rightarrow w U$ and $V \rightarrow
w V$, where $w$ is any diagonal matrix with $|w_{ii}|=1$, the operator
$\Delta_\mu$ remains invariant under these transformations. This
reparametrization freedom is exactly what leads to the apparent `gauge'
dependence of the results in the 
WKB-approach~\cite{ClineJoyceKainulainen-II}.
As we show below, only the diagonal elements of $\Delta_\mu$
contribute to the constraint and kinetic equations to order $\hbar$,
which then implies that our results are reparametrization invariant.
Finally, the new mass operators in the diagonal basis are given by
\beqa
{\hat M_{0d}} \! &=& \! \phantom{-}\frac12 \left(
            U \hat M V^\dagger + V \hat M^\dagger U^\dagger \right)
\\
{\hat M_{5d}} \! &=& \! - \frac{i}{2} \left(
        U \hat M V^\dagger - V \hat M^\dagger U^\dagger \right).
\eeqa
The constraint and kinetic equations now correspond to the {\em hermitean}
and {\em antihermitean} parts of (\ref{Neq-sigma1}-\ref{Neq-sigma4}),
respectively. Because of the matrix structure, these equations contain
a number of commutator and anticommutator terms involving $g^s_a$ and the
various matrix-operators.  However, we shall now argue that
(\ref{Neq-sigma1}-\ref{Neq-sigma4}) can effectively be taken to be diagonal
to the order at which we are working.
Indeed, in the propagating basis (\ref{rotation}) the off-diagonal terms
are obviously suppressed by $\hbar$ when compared to the diagonal elements.
On the other hand, they appear in the diagonal equations through the
commutators $\hbar^{-1}[\hbar\Sigma_z,g_{ad}^s]
\equiv[\Sigma_z,g_{ad}^s]$ and $[\Delta_z,g_{ad}^s]$, and thus at the same
order as the diagonal elements.  This then immediately implies that, when the
dynamics of CP-violating densities is considered, the off-diagonals contribute
at second order in $\hbar$ in the diagonal equations and can be consistently
neglected.  With this it is now straightforward to show that, to first order
accuracy in $\hbar$, the constraint equations reduce to the following equations
for the {\it diagonal} entries of $g^s_{ad}$:
\beqa
&&
k_0 g^{s}_{0d} \;-\; sk_z g^{s}_{3d}
   \;-\; m_R g^{s}_{1d} \;-\; m_I g^{s}_{2d} = 0
\label{CEAN} \\
&&
  k_0 g^{s}_{3d} \;-\; sk_z g^{s}_{0d}
         \;+\; \frac12 \tilde m_R' \partial_{k_z} g^{s}_{2d}
         \;-\; \frac12 \tilde m_I' \partial_{k_z} g^{s}_{1d} = 0
\label{CEBN} \\
&&
  k_0 g^{s}_{1d} + \frac{s}{2} \partial_z g^{s}_{2d}
         + s \Delta_{zd} g^s_{1d} - m_R g^{s}_{0d}
         + \frac12 \tilde m_I' \partial_{k_z} g^{s}_{3d} = 0
\label{CECN} \\
&&
  k_0 g^{s}_{2d} - \frac{s}{2} \partial_z g^{s}_{1d}
         + s \Delta_{zd} g^s_{2d} - m_I g^{s}_{0d}
         - \frac12 \tilde m_R' \partial_{k_z} g^{s}_{3d} = 0,
\label{CEDN}
\eeqa
where $m_{R,I}$ are the real and imaginary parts of the eigenvalues
of $M_d$, $\Delta_{zd}$ is the diagonal part of $\Delta_z$ and
$\tilde m_R' \equiv m_R' - 2m_I\Delta_{zd}$ and
$\tilde m_I' \equiv m_I' + 2m_R\Delta_{zd}$.   Similarly, the kinetic
equation for $g^{s}_{0d}$ to first order in $\hbar$ becomes
\beq
      \partial_t g^{s}_{0d} + s\partial_z g^{s}_{3d}
            - \tilde m_R' \partial_{k_z}g^{s}_{1d}
            - \tilde m_I' \partial_{k_z}g^{s}_{2d} = 0.
\label{KEAN}
\eeq
Following our treatment in section 2, it is now straightforward
to eliminate $g^s_{ad}$ ($a=1,2,3$) from
equations~(\ref{CEAN}) and (\ref{KEAN}) to obtain the constraint equation
\beq
\left( k_0^2 - k_z^2 - |M_d|^2
         + \frac{s}{k_0}|M_d|^2 \Theta' \right) g^s_{0d} = 0
\label{dispersion2}
\eeq
and the kinetic equation
\beq
        k_0 \del_t g^s_{0d}
      + k_z \del_z g^s_{0d}
       -  \Big( \, \frac{1}{2}{|M_d|^2}'
              - \frac{s}{2k_0} (|M_d|^2 \Theta')' \, \Big)
            \partial_{k_z} g^s_{0d} = 0
\label{Nboltzmann1}
\eeq
for the number density function $g^s_{0d}$ alone. Eqs.~(\ref{dispersion2})
and (\ref{Nboltzmann1}) are  analogous to the one field
equations~(\ref{constraint-g0}) and (\ref{boltzmann1}).
The difference is that here we have $N$ different equations
corresponding to $N$ diagonal elements of $g^s_{0d}$ in the mass eigenbasis.
The derivative of the effective angle
$\Theta'$ appearing in (\ref{dispersion2}-\ref{Nboltzmann1}) is defined as
\beq
\Theta' = \Theta_d^{\,\prime}+2\Delta_{zd}
\label{theff}
\eeq
where the angles $\Theta_d$ are the complex phases of the elements 
in the diagonal mass matrix: $M_d \equiv |M_d| e^{i\Theta_d}$.
Since the mixing field contribution shows up as a shift in 
the derivative of the pseudoscalar phase $2\Delta_{zd}$, it implies 
that equations~(\ref{dispersion2}-\ref{Nboltzmann1}) are manifestly 
reparametrization invariant. We can convert (\ref{theff}) to an 
alternative form in terms of the original mass matrix and the 
rotation matrix $U$:
\beq
     |M_d|^2 \Theta'
            = - \frac 12\mbox{Im}\Big( U \big( M{M'}^\dagger
                                - M'M^\dagger \big) U^\dagger \Big)_d \,,
\label{trans1}
\eeq
which is also manifestly reparametrization invariant. This expression 
will be convenient for discussion of the chargino sector of the MSSM 
and the NMSSM in sections~\ref{MSSM} and~\ref{NMSSM}.

The final steps in going from equations (\ref{dispersion2}) and
(\ref{Nboltzmann1}) to kinetic equations for the mass eigenmodes are
exactly analogous to the single fermionic field case: each mass eigenmode
$i$ gets projected to its own energy shell $\omega_{si\pm}$ given by
(\ref{dispersion2}), and the corresponding spectral decomposition density
function $f_{si\pm}$ obeys a semiclassical Boltzmann equation identical
to (\ref{boltzmann1-sc})
\begin{equation}
        \del_t f_{si\pm}
      + v_{si\pm} \del_z f_{si\pm}
      + F_{si\pm} \partial_{k_z} f_{si\pm} = 0
\,,
\label{boltzeqn-2}
\end{equation}
with the corresponding group velocity
\beq
v_{si\pm} = \frac{k_z}{\omega_{si\pm}}
\,
\label{vel-i}
\eeq
and the CP-violating semiclassical force
\beq
F_{si\pm}  =  - \frac{{|M_{i}|^2}' }{2\omega_{si\pm}}
                 \pm \frac{s(|M_{i}|^2\Theta_i')'}{2{\omega_{0i}}^2}
\,
\label{scforce2}
\eeq
computed from expression (\ref{theff}) or (\ref{trans1}).  It has been
shown in~\cite{JPT-letter,JPT-thick,ClineJoyceKainulainen-II},
that the spin-dependent term in $F_{si\pm}$ gives rise to a CP-violating
source proportional to $|M_{i}|^2\Theta_i'$ in the diffusion equations. We
can therefore loosely call this factor the `source', and proceed to compute
it in some special cases.

\subsection{Charginos in the MSSM}
\label{MSSM}

We first compute the source in the transport equations for charginos in
the MSSM.  The chargino mass term reads
\begin{equation}
{\overline\Psi_R}\, M\, \Psi_L + {\rm h.c.}
\,,
\label{chargino-mass}
\end{equation}
where $\Psi_R = ({\tilde W}_R^+,{\tilde h}^+_{1,R})^T$
and $\Psi_L = ({\tilde W}_L^+,{\tilde h}^+_{2,L})^T$
are the chiral fields in the basis of winos.
The mass matrix reads
\begin{equation}
        M = \left( \begin{array}{cc} m_2  & gH_2^* \\
                                     gH_1^* & \mu     \end{array} \right)
\,,
\label{CharginoMatrix}
\end{equation}
where $H_1$ and $H_2$ are the Higgs field vacuum expectation values and
$\mu$ and $m_2$ are the soft supersymmetry breaking parameters. For a realistic
choice of parameters there is no spontaneous CP-violation in the MSSM, so to a
good approximation we can take the Higgs {\it vev's} to be
real~\cite{HuberJohnLaineSchmidt,HuberJohnSchmidt-I}.
The matrix $U$ in~(\ref{rotation}) can be parametrized
as~\cite{ClineJoyceKainulainen-II}
\begin{equation}
        U = \frac{\sqrt{2}}{\sqrt{\Lambda(\Lambda+\Delta)}}
         \left( \begin{array}{cc} \frac{1}{2}(\Lambda+\Delta) & a
\\
          -a^*             & \frac{1}{2}(\Lambda+\Delta)   \end{array} \right)
\label{v}
\end{equation}
with
\begin{eqnarray}
     a &=& g(m_2H_1 + \mu^*H_2^*)  \\
     \Delta &=& |m_2|^2 - |\mu|^2  + g^2 ( h_2^2 - h_1^2 ) \\
     \Lambda &=& \sqrt{\Delta^2 + 4|a|^2}
\,,
\label{a-Delta-Lambda}
\end{eqnarray}
where $h_i \equiv |H_i|$ are normalized such that the tree level $W$-boson
mass is $M_W^2 = g^2h^2/2$, $h^2 = h_1^2 + h_2^2$. 
The physical chargino mass eigenvalues are given by
\beq
     m_\pm^2 = \frac{1}{2}\left(|m_2|^2 + |\mu |^2 + g^2h^2\right)
     \pm \frac{\Lambda}{2}\,.
\label{charginomasses2}
\eeq
Upon inserting (\ref{CharginoMatrix}) and (\ref{v}) into (\ref{trans1})
it is straightforward to show that the source term for charginos becomes
\beq
    m^2_\pm \Theta_\pm'  =
         \mp  \frac{g^2}{\Lambda}\Im (\mu m_2) ( h_1h_2 )'
\,,
\label{chargino-source}
\eeq
where $\Theta_+$ ($\Theta_-$)  corresponds to the higgsino-like state when
$|\mu| > |m_2|$ ($|\mu| < |m_2|$). CP violation is here mediated via the
parameters $\mu$, $m_2$ and may in fact be large~\cite{PilaftsisWagner}.
The result~(\ref{chargino-source}) is in perfect agreement with the chargino
source obtained by a WKB method in Ref.~\cite{ClineJoyceKainulainen-II}.

\subsection{Charginos in the NMSSM}
\label{NMSSM}

In the NMSSM there is an additional singlet field $S$ in the Higgs sector.
The singlet field couples to higgsinos, and hence the higgsino-higgsino
component in the chargino mass matrix (\ref{CharginoMatrix}) is
generalized in the NMSSM:
\beq
    \mu \rightarrow  \tilde \mu \equiv \mu + \lambda{S},
\label{tildemu}
\eeq
where $\lambda$ is the coupling for higgs(ino)-higgs(ino)-singlet interaction.
Another consequence of this extension is the possibility to have spontaneous
transitional CP-violation~\cite{HuberJohnLaineSchmidt}, so the Higgs fields
$H_i$ are in general complex. When the parameters $a$, $\Delta$ and $\Lambda$ 
are defined as in equation (\ref{a-Delta-Lambda}) with $\mu \rightarrow 
\tilde \mu$, the matrix $U$ in equation~(\ref{v}) still diagonalizes 
$MM^\dagger$.

In the NMSSM we must account for the complex dynamical phases of the 
higgs fields,
and hence we have to be more careful with our definition of the higgs doublets.
Our choice of writing the mass matrices~(\ref{CharginoMatrix})
corresponds to parametrizing the higgs doublets $\Phi_i$ as
\cite{HaberKane}:
\beq
     \Phi_1 =
     \left( \begin{array}{l}
               h_1 e^{i\theta_1} \\   \phantom{-}h_1^-
            \end{array} \right), \quad
     \Phi_2 =
     \left( \begin{array}{l}
              \phantom{-} h_2^+ \\ h_2 e^{i\theta_2}
            \end{array}\right),
\label{Hparam}
\eeq
where $h^\pm_i$ are the charged higgs fields. Only one of the higgs phases
$\theta_i$ is physical, while the other gets eaten by the gauge fields in the
unitary gauge. We wish to choose the physical phase in such a way that
the corresponding field does not couple to the neutral weak boson.
Given the parametrization (\ref{Hparam}), this condition implies that
\begin{equation}
        h^2_1\theta^{\,\prime}_1 = h^2_2\theta^{\,\prime}_2.
\label{gaugec}
\end{equation}
Using the gauge constraint (\ref{gaugec}) we can write
\begin{equation}
        \theta^{\,\prime}_1 = \frac{h^2_2}{h^2}\theta^{\,\prime}
\,,\qquad
        \theta^{\,\prime}_2 = \frac{h^2_1}{h^2}\theta^{\,\prime}
\,,
\label{thetas}
\end{equation}
where $h^2=h_1^2 + h_2^2$, and $\theta = \theta_1 +\theta_2$ is the physical
CP-violating phase.

To get the explicit form for the CP-violating term, we insert the NMSSM mass
matrix~(\ref{CharginoMatrix}) into~(\ref{trans1}). After some algebra
one finds the following three terms giving rise to CP-violating sources in
the NMSSM:
\begin{equation}
      \Theta_{\rm NMSSM}^\prime = \Theta_{h_1h_2}^\prime
                                 + \Theta_\theta^\prime + \Theta_S^\prime
\,,
\label{NMSSM-sources}
\end{equation}
The first term is the following generalization of the chargino
source~(\ref{chargino-source}):
\begin{equation}
    m^2_\pm \Theta_{h_1h_2\pm}' =
         \mp \frac{g^2}{\Lambda}\Im \big( \tilde \mu m_2 e^{i\theta} \big)
           \left(h_1h_2\right)^{\,\prime}
\,
\label{S-h1h2}
\end{equation}
for the case involving a new scalar field $S$ and possibly complex higgs
fields. However, there are two new types of terms in the NMSSM.  The term
$\Theta_\theta^\prime$ is proportional to a derivative of the CP-violating
phase $\theta$ in the Higgs sector, and reads
\begin{equation}
     m^2_\pm \Theta_{\theta\,\pm}^{\prime} = - \frac{g^2 \theta '}{\Lambda}
                \left(
                   \Big( \Lambda \pm (|m_2|^2+|\tilde\mu|^2) \Big)
                  \frac{h_1^2 h_2^2}{h^2}
               \mp\Re\big( \tilde\mu m_2 e^{i\theta}\big)
                    h_1 h_2 \right)
.
\label{S-theta}
\end{equation}
Finally, the source $\Theta_S^\prime$ can be written as a derivative of
the singlet condensate:
\begin{equation}
    m^2_\pm \Theta_{S\pm}'  = \pm \frac{\lambda g^2}{\Lambda}
                        \Im\big(m_2 H_1 H_2 S^{\,\prime}\big)
   +\frac{\lambda g^2}{2\Lambda}\Big(\Lambda 
        \pm (|\tilde\mu|^2 + g^2 h^2 -  |m_2|^2)\Big)
                  \Im(\tilde\mu^* S')
 \,.
\label{S-S}
\end{equation}
In all formulae (\ref{S-h1h2}-\ref{S-S}) the mass eigenvalues $m_\pm^2$
can be read off from equation (\ref{charginomasses2}) with the
replacement $\mu \rightarrow \tilde \mu$, where $\tilde \mu$ is given by
Eq.~(\ref{tildemu}). Baryogenesis in the NMSSM from the semiclassical
force has been studied in Ref.~\cite{HuberJohnSchmidt}.

\section{Mixing bosonic fields}
\label{Mixing bosonic fields}

Here we first show that, unlike for fermions, the constraint and kinetic
equations for mixing bosons acquire no gradient correction to first order
in $\hbar$ in the collisionless limit. We then derive the constraint
and kinetic equations accurate to second order in gradients which can be
used as a starting point for baryogenesis calculations.
$N$ mixing bosonic fields with a spatially varying mass matrix $M^2$
obey the Klein-Gordon equation
\begin{equation}
\big( \Box_u + M^2(u) \big) \phi(u) = 0,
\label{KGeq}
\end{equation}
where $\phi$ is an $N$-dimensional vector whose components are coupled by
the hermitean mass matrix $M^2$.
Multiplying (\ref{KGeq}) from the left by $-i\phi^\dagger(v)$ and
taking the expectation value with respect to the initial state we get
\begin{equation}
     \big( \Box_u + M^2(u) \big) G^<(u,v) = 0.
\label{SDboso}
\end{equation}
where the Wightman function $G^<$ is defined as
\begin{equation}
G^<(u,v) = -i\langle \phi^\dagger(v)\phi(u) \rangle.
\end{equation}
After performing the Wigner transform, equation~(\ref{SDboso}) becomes
\begin{equation}
    \left( \frac14 \del^2 - k^2 - i  k\cdot\del + M^2
  e^{-\frac i2\stackrel{\leftarrow}{\del}\,\cdot\,\del_k}
  \right) G^< = 0.
\label{bosons-eom}
\end{equation}
In the case when $N=1$ it is immediately clear that the first quantum
correction to the constraint equation
(the real part of (\ref{bosons-eom})) is of
second order and to the kinetic equation (imaginary part) of third order in
gradients (second order in $\hbar$).
To extract the spectral information to second order in
$\hbar$ is quite delicate since the constraint equation
in~(\ref{bosons-eom}) contains derivatives~\cite{BKP}.
In the case of more than one mixing fields it is convenient to
rotate into the mass eigenbasis, just as in the fermionic case:
\begin{equation}
M_d^2 = U M^2U^\dagger ,
\label{rotu}
\end{equation}
where $U$ is a unitary matrix. In the propagating basis the
equation~(\ref{bosons-eom}) becomes
\begin{equation}
    \left( \frac14 {\cal D}^2 - k^2
      - i  k\cdot{\cal D}
      + M^2e^{-\frac i2\stackrel{\leftarrow}{{\cal D}}\,\cdot\,\del_k}
     \right) G_d^< = 0,
\label{bosons-eom-d}
\end{equation}
where $G^<_d \equiv U G^< U^\dagger$ and
the `covariant' derivative is defined as:
\def\calD{{\cal{D}}}
\begin{equation}
\calD_\mu = \partial_\mu -i \left[{\Xi}_\mu,\;\cdot\;\right],
\qquad
{\Xi}_\mu = iU\del_\mu U^\dagger
.
\label{calD}
\end{equation}
Since $(G_d^<)^\dagger= -G_d^<$ and
$\calD_\mu^\dagger = \calD_\mu$, we identify the antihermitean part
of~(\ref{bosons-eom-d}) as the constraint equation:
\begin{equation}
- 2k^2 G_d^<  + \left\{ \hat M^2_c +\frac14 {\cal D}^2 , G_d^< \right\}
              -i \left[  k\cdot{\cal D} + \hat M^2_s,
                G_d^< \right]
              = 0,
\label{bosons-ce-d}
\end{equation}
and the hermitean part is the kinetic equation
\begin{equation}
     \left\{  k\cdot{\cal D} + \hat M^2_s,G_d^< \right\}
     -i  \left[ \hat M^2_c +\frac14 {\cal D}^2 , G_d^< \right]
     = 0,
\label{bosons-ke-d}
\end{equation}
where we defined
\begin{eqnarray}
\hat M^2_c &=& M^2_d\cos\frac 12\stackrel{\leftarrow}{{\cal D}}\cdot\,\del_k
\nn\\
\hat M^2_s &=& M^2_d\sin\frac 12\stackrel{\leftarrow}{{\cal D}}\cdot\,\del_k
.
\label{McMs}
\end{eqnarray}
We now use the analogous argument as in the fermionic case in section 3.
The off-diagonal elements of $G^<_d$ in (\ref{bosons-ce-d}-\ref{bosons-ke-d})
are sourced by the diagonal elements
through the terms involving commutators which are suppressed by at least
$\hbar$ with respect to the diagonal elements. This implies
that, in order to capture the leading order nontrivial effect in gradients,
we can work in the diagonal (semiclassical) approximation for $G^<_d$.
By inspection of (\ref{bosons-ce-d}-\ref{bosons-ke-d}) we can
now immediately write the constraint and kinetic equations in the
diagonal approximation accurate to order $\hbar$ as follows
\beqa
\left(k^2 - M_d^2 \right) G^<_d &=& 0 \\
\left( k\cdot\del + \frac12 (\partial M_d^2) \cdot
       \,\partial_{k} \right) \,G^<_d &=& 0.
\label{lastihope}
\eeqa
In contrast to the fermionic equivalent (\ref{dispersion2}),
these equations contain no CP-violating corrections
to order $\hbar$ and display only the usual classical CP-conserving term
associated with the mass eigenvalues.
This analysis is relevant for example for calculation of the CP-violating
force in the  {\it stop} sector $\tilde q = (\tilde t_L,\tilde t_R)^T$
of the MSSM, in which the mass matrix reads
\begin{equation}
       M_{\tilde q}^2 = \left(
             \begin{array}{cc} m_Q^2          & y( A^* H_2 + \mu H_1)   \\
                        y( A H_2 + \mu^* H_1) & m_U^2 \end{array} \right)
\,,
\label{squarks}
\end{equation}
where $m_{Q}^2$ and $m_U^2$ denote the sum of the soft susy-breaking
masses, including D-terms and $m_t^2 = y^2 H_2^2$. Our analysis
immediately implies that for squarks there is no CP-violating
correction to the dispersion relation at first order in gradients,
and hence there is no CP-violating semiclassical force in
the kinetic equation at order $\hbar$.

\section{Continuity equations and CP-violating sources}
\label{Cont-sources}

The quantities eventually relevant for baryogenesis are CP-violating
fluxes.  We now investigate how the CP-violating sources appear in the
equations for the divergences of the vector and axial vector currents.
Let us first consider the divergence of the vector current. We have
\beqa
\partial_\mu j^\mu &=& \partial_\mu
        \langle \bar\psi(x) \gamma^\mu \psi(x) \rangle \\
        &=& \int \frac{d^{\,2}k}{(2\pi)^2} \partial_\mu
             {\rm Tr}(-iG^< \gamma^\mu)
\label{vecdiv1}
\eeqa
where the derivative in the integral expression is taken with respect
to the center-of-mass coordinate. Using the decomposition (\ref{glesss})
and the constraint equations to write $g^s_{3i}$ in terms of $g^s_{0i}$
($i$ is the flavour index), Eq.~(\ref{glesss}) is easily shown to reduce
to just a momentum integral over the kinetic equation~(\ref{Nboltzmann1}).
Hence we get the usual continuity equation
\beq
  \partial_\mu j_{si\pm}^\mu \equiv
   \del_t n_{si\pm} + \partial_z(n_{si\pm} u_{si\pm}) = 0,
\label{Ncont}
\eeq
showing that the vector current is conserved, and contains no sources.
The fluid density $n_{si\pm}$ and velocity 
$u_{si\pm}\equiv \langle v_{si\pm} \rangle$ are defined as
\begin{eqnarray}
n_{si+}  &\equiv& \int_+\frac{d^{\,2}k}{(2\pi)^2} g^{s}_{0i}
\nonumber\\
n_{si+} \langle v_{si+}^p \rangle &\equiv&
        \int_{+}\frac{d^{\,2}k}{(2\pi)^2}
       \left( \frac{k_z}{k_0} \right)^p g^{s}_{0i} \,,
\label{Ncont-2}
\end{eqnarray}
where $\int_+ \equiv \int_{k_0\geq 0}$ denotes integration over 
the positive frequencies. The fluid density $n_{si\pm}$ should not
be confused with the phase space density $n_s$ in~(\ref{fs}).
To get the density and velocity moments for
antiparticles one should integrate over the negative frequencies and make
use of~(\ref{fs}). Here we have again restricted
ourselves to 1+1 dimensions; in 3+1 dimensions the expressions differ
in detail, but not in essence~\cite{KPSW2}. No source appears
in~(\ref{Ncont}) simply because the semiclassical force term
in~(\ref{Nboltzmann1}) reduces to vanishing boundary terms at
$k_z\rightarrow\pm\infty$.  In the case of a single Dirac fermion this
result remains valid to any order in gradients. In a general case of
mixing fields the dynamics of off-diagonal
elements of $g^s_{0}$ may induce sources, which would however be higher
than first order in $\hbar$. Our proof that there are no CP-violating
sources to the continuity equation for the vector current is contrary to
the results of Refs.~\cite{Riotto-I,Riotto-II}.  Our derivation is more
general than that of~\cite{Riotto-I} in that by treating mass as part of
the flow term it includes the ``mass resummation" of
Refs.~\cite{Riotto-I,Riotto-II} to infinite order,
it is not based on any particular {\it Ansatz} for $g_{0d}^s$, and finally,
but most importantly, we took correct account of the constraint equations.
The fact that we have not treated collisions terms here does not resolve
the differences, because the collisional contributions arise only from
nonsingular loop diagrams which were not treated in Ref.~\cite{Riotto-I}
either.

We next consider the continuity equation for the axial vector current.
This can be obtained from Eq.~(\ref{Ncont}) by replacing $\gamma^\mu$ by
$\gamma^\mu\gamma^5$. The presence of $\gamma^5$ essentially changes the
roles of $g^s_3$ and $g^s_0$, so that the axial divergence reduces to an
integral over the kinetic equation for $g^s_3$, which can be inferred
from~(\ref{Neq-sigma2}):
\begin{equation}
     \partial_t g^s_3 + s\partial_z g^s_0
   + 2\left(M_I - \frac 18 M_I^{\prime\prime}\partial_{k_z}^2\right) g^s_1
   - 2\left(M_R - \frac 18 M_R^{\prime\prime}\partial_{k_z}^2\right) g^s_2
   = 0.
\label{Neq-sigma2b}
\end{equation}
It is thus easy to see that the divergence of the axial current acquires
the expected form:
\begin{equation}
      \partial_\mu j^\mu_{5s} = -2iM_R \langle\bar{\psi}_s\gamma^5\psi_s\rangle
                             - 2M_I \langle\bar{\psi}_s\psi_s\rangle,
\label{Neq-sigma2c}
\end{equation}
where $\langle\bar{\psi}_s\psi_s\rangle = \int_k g^s_1$ denotes the scalar,
and $\langle\bar{\psi}_s\gamma^5\psi_s\rangle = i \int_k g^s_2 $ the
pseudoscalar density, and $\int_k \equiv \int d^2k/(2\pi)^2$.
We now make use of the constraint equations~(\ref{Neq-sigma1}-\ref{Neq-sigma4})
to express $g^s_1$, $g^s_2$ and $g^s_3$ in~(\ref{Neq-sigma2b}) in terms of
$g^s_0$ (to second order in gradients), diagonalize~(\ref{Neq-sigma2b}) by
rotating  into the propagating basis, where we can take $g^s_{0d}$ to be
diagonal to order $\hbar$ accuracy, and finally integrate over the momenta.
We thus arrive at the following equation for the axial current divergence:
\beqa
  \partial_t\left(n_{si\pm} u_{si\pm} \right)
         + \del_z n_{si\pm}
    \!\! &-& \!\! \left(
           \vert M_{i}\vert^{2}\partial_z
         + \frac 12 \vert M_{i}\vert^{2\,\prime}
         \right) 
{\cal I}_{2si\pm}
\nn \\ 
    \!\!  &\pm& \!\! s\left(
       \vert M_{i}\vert^{2}\Theta_i'\partial_z
       + \frac 12 (\vert M_{i}\vert^{2}\Theta_i')'
       \right){\cal I}_{3si\pm}
       = 0,
\label{cont-axial-I}
\eeqa
where we defined
\beq
{\cal I}_{psi+} = \int_{+}\frac{d^2 k}{(2\pi)^2} \frac{g_{0i}^s}{k_0^p}
\; = \; 
                  \int \frac{dk_z}{8\pi Z_{si\pm}}
                   \frac{f_{si\pm}}{\omega_{si\pm}^p}
\quad (p=2,3),
\label{iota}
\eeq
As usual, one has to introduce some truncation scheme to close
the equations to two unknown quantities (here $n_{si\pm}$ and
$u_{si\pm}$). There is of course some freedom
as to how to do this step, and one might implement the truncation
in (\ref{cont-axial-I}) by replacing the distributions $f_{si\pm}$ 
in the ${\cal I}_{psi\pm}$-integrals by the equilibrium ones. However, 
it is instructive to observe that, by extracting a total derivative from
the $\vert M_{i}\vert^{2}$-terms in (\ref{cont-axial-I}), the corresponding
integrals can be combined with the $\partial_z n_{si\pm}$-term
 to give a second velocity moment term,
in terms of which (\ref{cont-axial-I}) becomes simply
\begin{equation}
   \partial_t \left(n_{si\pm} u_{si\pm}  \right)
     + \del_z \left(n_{si\pm} \langle v_{si\pm}^2 \rangle \right)
     = S_{si\pm},
\label{cont-axial}
\end{equation}
where the source $S_{si\pm}$ is given by the average over the
semiclassical force (\ref{scforce2}) divided by $k_0$:
\beq
    S_{si\pm} =
    - \frac 12 \vert M_{i}\vert^{2\,\prime} {\cal I}_{2si\pm}
    \pm \frac 12 s (\vert M_i\vert^{2}\Theta_i')'  {\cal I}_{3si\pm}.
\label{realsource}
\eeq
This equation can be truncated by the method standardly used for moment
expansion. One writes $\langle v^2_{si\pm} \rangle
\rightarrow u_{si\pm} ^2 + \sigma^2_{si\pm}$ and uses the
equilibrium distributions for $f_{si\pm}$'s when evaluating the
variance $\sigma^2_{si\pm} \equiv \langle v_{si\pm}^2 \rangle
- u_{si\pm}^2$ and the remaining integrals
${\cal I}_{psi\pm}$ appearing in the source term (\ref{realsource}).

Remarkably, expressions (\ref{cont-axial}-\ref{realsource}) show that
the divergence of the axial current in fact corresponds to the first
velocity moment of the kinetic equation for $g^s_{0d}$. Because a
nontrivial CP-violation is tied to nontrivial complex
phases in the {\em pseudoscalar}, or axial mass term, the fact that the
source appears in the axial current nicely explains why the semiclassical
source appears at first order in moment expansion in
earlier semiclassical
treatments~\cite{JPT-letter,JPT-thick,ClineJoyceKainulainen-II}.

The first source in (\ref{realsource}) is to leading order in gradients
spin-independent and does not violate CP. 
It is important for the phase transition dynamics however, in that it 
provides the dominant contribution to the friction on
bubble walls from fermions~\cite{MooreProkopec95,JohnSchmidt01}. The
second source is spin-dependent and CP-violating and it is thus responsible
for baryogenesis.  This is one of the main results of this paper, as it
shows how the source from the semiclassical force enters to momentum
integrated transport equations used in practical calculations. To promote
equations ~(\ref{Ncont})  and~(\ref{cont-axial}) into transport equations
for baryogenesis calculations we still need to generalize them to include
collisions as indicated in~(\ref{KEfull}), which will be done
elsewhere.

\subsection{Spontaneous source in relaxation time approximation}

 We have so far shown how the source arising from the semiclassical
force enters in the equations for currents, or equivalently momentum
averaged transport equations. We shall now give a heuristic account
on how sources have been modeled elsewhere in literature. The method very
often used in EWBG considerations, apart from the WKB-computations,
employs the relaxation time approximation for the kinetic equations.
Here we discuss how the relaxation time approximation can be incorporated 
into our formalism, and then make a comparison with literature. 
Including collisions in the relaxation time approximation into 
the Liouville equation~(\ref{boltzeqn-2}) results in the following kinetic 
equation for the distribution function $f_q$ for a charge $q$: 
\begin{equation}
     \left(\partial_t + \vec v_q\cdot \partial_{\vec x}
         + \vec F_q\cdot \partial_{\vec k}\right) \! f_q
         = - \frac{ f_q - f_{q0}}
      {\tau_q}
\,,
\label{relaxation}
\end{equation}
where $\tau_q\equiv \Gamma_q^{-1}$ is the equilibration time for $q$
(which we expect to be given by the relevant elastic scattering rate),
$\vec F_q$ the semiclassical force and $f_{q0}$ is the thermal 
equilibrium distribution function. In presence of a background field 
that violates $q$, one expects $f_{q0}$ to be shifted with respect to 
the naive thermal equilibrium, leading to a `spontaneous' source that 
violates $q$. This source is more important for thick walls when the 
equilibrium $f_q\approx f_{q0}$ is approximately attained on the wall. 
The {\em spontaneous baryogenesis} source was originally introduced by 
Cohen, Kaplan and Nelson~\cite{CohenKaplanNelson91,HuetNelson-II} in
the context of two Higgs doublet models, and then subsequently refined
to include the $m^2$-suppression in~\cite{ComelliPietroniRiotto,JPT-thick}.
The derivation was successively reconsidered in~\cite{HuetNelson-I,
CarenaQuirosRiottoViljaWagner,CarenaMorenoQuirosSecoWagner}.
For example, in~\cite{CarenaMorenoQuirosSecoWagner} the CP-violating 
vector current $j^0_q = \int (dk_z/(2\pi)) f_{q0}$ for charginos in 
the MSSM was computed and inserted into the transport equations 
written in the relaxation-time approximation.

The spontaneous baryogenesis source can be in our formalism obtained simply
by integrating~(\ref{relaxation}) over the momenta. The source then becomes
the CP-violating contribution to the vector current~(\ref{vector2}), 
which to first order in gradients (or equivalently $\hbar$) reads
\begin{equation}
   n_{si} \equiv   j^0_{si+} -  j^0_{si-} \approx 
   s |M_i|^2\Theta_i'
   \int_{\omega\geq |M_i|} \frac{d\omega }{4\pi}\,
          \frac{1}{\sqrt{\omega^2-|M_i|^2}} \, 
   \frac{f_{\omega}}{\omega^2}
         \left(1+\frac{\omega}{T}(1-f_{\omega})\right),
\label{vector3}
\end{equation}
where we approximated  $f_{si\pm}$ by the equilibrium distribution 
function in plasma frame, 
$f_{si\pm} \rightarrow 1/(e^{\omega_{si\pm}/T}+1) \approx f_{\omega} 
\mp (s|M_i|^2\Theta_i'/2\omega^2)df_{\omega}/d\omega$, where 
$f_{\omega} = 1/(e^{\omega/T}+1)$, and we used $k_z dk_z=\omega d\omega$.
To make a comparison with literature, note first that the spontaneous
source~(115)  is {\it nonanalytic} in $|M_i|^2$. Since earlier 
attempts~\cite{HuetNelson-I,Riotto-I,Riotto-II,
CarenaQuirosRiottoViljaWagner} used expansions in powers of $|M_i|^2$ 
to compute the spontaneous source, their results are at best
incomplete.  Consider next the CP-violating source for charginos in 
the MSSM. According to our equation (\ref{chargino-source}) it is given 
by $|M_d|^2\Theta' =$ ${\rm diag}(m_+^2\Theta'_+,m_-^2\Theta'_-)$, where 
$m^2_\pm \Theta_\pm' = $ $\mp ({g^2}/{\Lambda})\Im (\mu m_2) ( h_1h_2)'$,
showing the parametrical dependence $ (h_1h_2)'$ on the higgs fields. 
This is in contrast with 
Refs.~\cite{CarenaQuirosRiottoViljaWagner,CarenaMorenoQuirosSecoWagner},
where a source proportional to $h_1h_2'-h_2h_1'$ was found and claimed 
to be important for baryogenesis. The origin of the difference may be 
in the fact that we made use of the constraint equations, which is
necessary to obtain the correct results.

Consider now the axial vector current. The corresponding spontaneous
source can be easily obtained from~(\ref{axial2}):
\begin{equation}
{s}j^0_{5ds\pm}  =
    \sum_{s_{k_z}=\pm} s_{k_z} \int \frac{d\omega }{8\pi}\,
                      f_{sd\pm} = 0,
\label{axial3}
\end{equation}
where we took $f_{si\pm}\rightarrow 1/(e^{\omega/T}+1)$. As a consequence,
there is {\it no} spontaneous baryogenesis source from the axial vector
current when computed in the relaxation time approximation.

An attempt to compute the spontaneous baryogenesis source was made 
by Riotto~\cite{Riotto-I}, where the divergence of the vector current 
was computed in the Schwinger-Keldysh formalism~\cite{Mahan-PhysReps} 
and then, based on~\cite{HuetNelson-I}, identified with a spontaneous 
source~\cite{Riotto-II}. In this way he found an equation which 
formally reads: $\partial_\mu j^\mu_q \sim {\rm spontaneous \; source}$.
According to Eq~(\ref{Ncont}) however {\em no} source appears in the 
continuity equation for the vector current.  Instead sources appear in 
the continuity equation for the {\em axial} vector 
current~(\ref{cont-axial-I}-\ref{cont-axial}), which has not been 
so far considered in literature.

\section{Discussion and summary}

The question of a first principle derivation of CP-violating fluxes in
transport equations has been the main theoretical challenge of recent
work on electroweak baryogenesis. In this paper we derive the kinetic
equations appropriate for EWBG in a systematic gradient expansion
starting from the (Dirac) equation of motion for the two-point Wightman
function $G^<$ in the collisionless limit. The gradient expansion we use
is well controlled and corresponds to an expansion in the de Broglie 
wave length divided by the wall width. In EWBG applications one typically 
has $\ell_{dB}/\ell_w \ll 1$, so that such an expansion should be rapidly
converging.

We have shown that to first order in $\hbar$ the collisionless kinetic
equations for both fermions and bosons can be recast as the Liouville
equations for a single particle distribution function where the group 
velocity and the semiclassical force terms contain all quantum 
information, and in particular the CP-violating terms which source 
baryogenesis. These results agree with 
Ref.~\cite{ClineJoyceKainulainen-II}, where the kinetic equations 
were obtained in the semiclassical WKB picture, originally developed 
for EWBG problem in~\cite{JPT-letter,JPT-thick}. The outstanding 
contribution of this paper is in a first principles derivation of 
these results in a completely controlled approximation scheme. We 
also derive the semiclassical force in the general case of $N$ mixing 
fermions and in particular for the chargino sector in both the MSSM 
and NMSSM. Finally we prove that there is no CP-violating force at 
first order in $\hbar$ for scalar fields~(\ref{lastihope}). Let us 
point out that the fact that the quasiparticle picture of plasma 
still holds to order $\hbar$ in gradient expansion is not 
surprising since the gradient correction for fermions from 
a pseudoscalar (CP-violating) mass condensate can be equivalently 
reformulated in terms of a `classical' axial vector field 
condensate~\cite{JPT-letter,JPT-thick}.

We have also studied the vector and axial vector current equations, and 
showed that, while the vector current is conserved, the axial current
contains both CP-conserving and CP-violating sources. This is to be 
expected, as CP-violation in particular is known to be caused by complex
pseudoscalar (axial) mass terms.  We have then pointed out that the axial
current equation corresponds to the first velocity moment of the kinetic 
equation. This explains why the CP-violating source appears at first 
order in moment expansion \cite{JPT-letter,JPT-thick}. We finally made 
connection between the present results and literature where the continuity
equations were written in the relaxation time approximation.
In this context  the source has been claimed to appear either 
as the vector current divergence~\cite{Riotto-I,Riotto-II} or the
time-component of the vector 
current $j^0$~\cite{HuetNelson-I,CarenaQuirosRiottoViljaWagner,
CarenaMorenoQuirosSecoWagner}. However, we have shown here that the vector
current continuity equation~(\ref{Ncont}) in fact contains no source. 
We have also computed $j^0$, and applied the result to the MSSM, and found
a parametrically different result
from Refs.~\cite{HuetNelson-I,CarenaQuirosRiottoViljaWagner,
CarenaMorenoQuirosSecoWagner}.

For simplicity here we consider only the collisionless limit in 1+1 
dimensions. One can show~\cite{KPSW2} that generalization to the 3+1
dimensional case does not affect our discussions in any qualitative 
way. The question of how to consistently include collisions 
we postpone to a future publication.

\section*{Acknowledgments}

KK and TP wish to acknowledge Dietrich B\"odeker
and Michael Joyce for useful discussions 
and for collaboration on related issues.

\end{document}